\documentclass[aps,prb,reprint,groupedaddress,showpacs]{revtex4-1}
\usepackage{amsmath}
\usepackage{amssymb}
\usepackage[colorlinks,bookmarks=false,citecolor=red,linkcolor=blue,urlcolor=blue]{hyperref}
\def\Hhat{\hat{H}}

\def\rhohat{\hat{\rho}}
\def\Xhat{\hat{X}}
\def\Phat{\hat{P}}

\def\That{\hat{T}}
\def\calXhat{\hat{\cal X}}
\def\Identity{\mathbb{I}}
\def\calU{\mathcal{U}}
\begin{document}
\title{Exact spin-orbital separation in a solvable model in one dimension}
\author{Brijesh Kumar}
\email{bkumar@mail.jnu.ac.in}
\affiliation{School of Physical Sciences, Jawaharlal Nehru University, New Delhi 110067, India}
\date{April 16, 2013} 
\begin{abstract} 
A one-dimensional model of coupled
spin-1/2 spins and pseudospin-1/2 orbitals with nearest-neighbor interaction is rigorously shown to exhibit spin-orbital separation by means of a non-local unitary transformation. On an open chain, this transformation completely decouples the spins from the orbitals  in such a way that the spins become paramagnetic while the orbitals form the soluble XXZ Heisenberg model. The nature of various correlations is discussed. The more general cases, which allow spin-orbital separation by the same method, are pointed out. A generalization for the orbital pseudospin greater than 1/2 is also discussed. Some qualitative connections are drawn with the recently  observed 
spin-orbital separation in Sr$_2$CuO$_3$.
\end{abstract}
\maketitle

 \section{Introduction} The multi-orbital Mott-Hubbard insulators are known to exhibit a rich variety of orbito-magnetic phases due to the coupled nature of spin and orbital degrees of freedom. The theoretical framework that one uses to discuss such problems goes by the name of Kugel-Khomskii models. These are natural extensions of the Heisenberg spin-exchange to the multi-orbital cases wherein the spins and orbitals get coupled via superexchange to form the spin-orbital models.~\cite{Kug.Kho.1,Ranninger} The phonons, through Jahn-Teller coupling, also participate in the systems with orbital degeneracy. However, at the very least, one simply focuses on the electronic part, the leading term in energy, that gives rise to the spin-orbital models within second order perturbation theory in the limit of strong local repulsion for effectively one electron or hole per unit cell.

There are a lot of materials, such as KCuF$_3$, V$_2$O$_3$, LaMnO$_3$, MnF$_3$, (Na,\,Li)NiO$_2$, Sr$_2$CuO$_3$ (to name a few), where the Kugel-Khomskii type spin-orbital models are directly applicable.~\cite{Kug.Kho.1,Ranninger, KCuF3,Mos.Kho,Vernay,nature.Sr2CuO3} For example, in KCuF$_3$, the Cu$^{2+}$ ions, having [Ar]$3d^9$ configuration in the octahedral crystal field of $F^-$ ions, can be treated as having one hole in the twofold degenerate $e_g$ orbitals ($d_{x^2-y^2}$ and $d_{z^2}$). It is a Mott insulator, and has been studied as a Kugel-Khomskii problem.~\cite{KCuF3.1,Mos.Kho} There are also cases of a different type in which the effective problem has the form of a spin-orbital model, although in actual one may not be dealing with the orbitals.~\cite{subrah,BeccaMila,Sa.Gro} For instance, an odd-legged spin-1/2 tube can be described in terms of the effective spin-1/2 and chirality variables. Since the chirality can be treated as a pseudospin-1/2 object, an effective model for such a spin-tube is a spin-orbital model in one dimension.~\cite{tube1,tube3,tube4} Clearly, there is much interest in studying these model problems with different motivations. 

In this paper, we present a solvable spin-orbital model in one dimension (1d), whose most significant feature is the `spin-orbital separation', an effect similar to the spin-charge separation in the interacting 1d electrons (Luttinger liquids). Our model has a realistic form. It resembles the effective models for odd-legged spin-1/2 tubes,~\cite{tube1, tube4} and may be motivated by a microscopic two-band Hubbard model.~\cite{Kug.Kho.1,Ranninger} This work is built upon an earlier brief suggestion of spin-orbital decoupling by the present author.~\cite{BK.09} Interestingly, the recent observation of spin-orbital separation in Sr$_2$CuO$_3$ is a welcome development that makes our independent theoretical findings experimentally relevant.~\cite{nature.Sr2CuO3}

In a spin-orbital 
model, the electron spin is described by the Pauli operators $\vec{\sigma}^{ }_l$, where $l$ is the site index of a lattice.  The operators for the orbitals depend upon the details thereof. Here, we consider the two orbital case, for which the operators in the orbital space can be treated as pseudospin-1/2 objects described by another set of Pauli operators, $\vec{\tau}^{ }_l$. For a system of spin-1/2 spins and pseudospin-1/2 orbitals, a coupled spin-orbital problem can have the following generic form.~\cite{Pat.Sin.Kho,Kug.Kho.2}
\begin{eqnarray}
\Hhat^{ }_{so} = \sum_l\left[J_1 \vec{\sigma}_l\cdot\vec{\sigma}_{l+1} +  J_2\left\{\vec{\tau}_l,\vec{\tau}_{l+1}\right\}_{\Delta} \right. \nonumber \\
 \left. + J_3 \left(\vec{\sigma}_l\cdot\vec{\sigma}_{l+1}\right)\left\{\vec{\tau}_l,\vec{\tau}_{l+1}\right\}_{\Delta^\prime}\right] \label{eq:H_KK}
\end{eqnarray}
Here, the model Hamiltonian, $\Hhat^{ }_{so}$, is written on a 1d lattice, and it only has the nearest-neighbor exchange interactions. Of course, in general, the $\Hhat^{ }_{so}$ could have further neighbor exchanges, and be living on any lattice. The exchange interactions, $J_{\{1,2,3\}}$, are assumed to be antiferromagnetic, unless stated to be otherwise. The symbol $\{\vec{\tau}_l,\vec{\tau}_{l+1}\}_\Delta$ denotes the anisotropic exchange, $\Delta\tau^z_l\tau^z_{l+1}+2\left(\tau^+_l\tau^-_{l+1}+\tau^+_{l+1}\tau^-_{l}\right)$, between the orbitals. Evidently, the $\Hhat^{ }_{so}$ is SU(2) symmetric for spins, and has U(1) symmetry for orbitals. Therefore, the total spin, and the $z$-component of total pseudospin, are conserved. When $\Delta=\Delta^\prime=1$, it becomes SU(2)$\times$SU(2) symmetric. These are complex models to investigate theoretically, and have attracted a lot of attention.

Of the many cases of $\Hhat^{ }_{so}$, quite a few are special as they allow for exact analytical solution of some kind. The most notable is the SU(4) symmetric model corresponding to $J_1=J_2=J_3$ and $\Delta=\Delta^\prime=1$, whose ground state energy and elementary excitations are known using Bethe ansatz.~\cite{Yam.Shi.Ued,Sutherland.SU4} There are other interesting cases as well. For example, when $J_1=J_2=3J_3$ for $\Delta=\Delta^\prime=1$, the exact ground state is doubly degenerate with the spins and the orbitals separately forming pairwise singlets on alternate bonds. The same ground state also holds true for $\Delta=\Delta^\prime=0$ and $J_1=2 J_3$ and $J_2=3J_3$.~\cite{Kol.Mik, Itoh, Kol.Mik.Sho} The  quantum phase diagram for general $J_1/J_3$ and $J_2/J_3$ has also been investigated using the density matrix renormalization group (DMRG) method.~\cite{Pati.Singh,Pat.Sin.Kho} Different variations of $\Hhat_{so}$ are also known to give some interesting solvable models. Notable ones are a symmetric XY spin-orbital model which can be completely solved using Jordan-Wigner transformation,~\cite{Mila.PRL99} and an integrable 
dimerized chain with $J_3=-2$ and $(J_1,J_2)$ alternately equal to $(\frac{3}{2},\frac{1}{2})$ or $(\frac{1}{2},\frac{3}{2})$ on successive nearest-neighbor bonds.~\cite{Mar.Nie} We too have found a special case which is the subject of our discussion in this paper.

\section{Model} We present the following spin-orbital model in 1d. We were led to this special model unexpectedly while working on the infinite $U$ Hubbard model.~\cite{BK.09, BK.08}
\begin{equation}
\Hhat= \sum_l \left[J_z \tau^z_l\tau^z_{l+1} + J_\perp \Xhat^{ }_{l,l+1}\left(\tau^+_l\tau^-_{l+1} + \tau^+_{l+1}\tau^-_l \right)\right]
\label{eq:H}
\end{equation}
Here, $J_z$ and $J_\perp$ are the nearest-neighbor interaction parameters. While $J_z$ measures the strength of Ising interaction between orbitals, $J_\perp$ couples the spin-exchange, $\Xhat^{ }_{l,l+1} = (1+\vec{\sigma}^{ }_l\cdot\vec{\sigma}^{ }_{l+1})/2$, with the orbital quantum fluctuations (the XY part of the orbital-exchange). This model corresponds to the $\Hhat_{so}$ for $J_1=\Delta^\prime=0$, $J_2=J_3=J_\perp/4$ and $\Delta=4J_z/J_\perp$. Evidently, the $\Hhat$ is SU(2)$\times$U(1) symmetric, and conserves $\sum_l\tau^z_l$ as well as the total spin, $\sum_l\vec{\sigma}^{ }_l$. Below we show that the $\Hhat$ of Eq.~(\ref{eq:H}) is exactly soluble through a unitary transformation which completely decouples the spins from the orbitals, and the decoupled problems of spins and orbitals are separately solvable. Recently,  there has been some experimental interest in the spin-orbital separation in quasi 1d systems.~\cite{nature.Sr2CuO3} The model $\Hhat$ presents a realistic and rigorous theoretical example of the same. 

\subsection{Spin-orbital separation}
An important property of the spin-exchange operators, $\Xhat^{ }_{l,l+1}$, is that $\Xhat^2_{l,l+1} = \Identity$, where $\Identity$ is the Identity operator. It prompts us to view $\Xhat^{ }_{l,l+1}$ as a unimodular `operator phase' factor. Since it appears in conjunction with $\tau^+_l\tau^-_{l+1}$ in $\Hhat$, we are led to think that one may suitably gauge-transform $\tau^\pm_l$ to rid the $\Hhat$ of $\Xhat^{ }_{l,l+1}$ operators. It turns out that we can indeed do this by the procedure invented in Ref.~\onlinecite{BK.09} by the present author for the infinite-$U$ Hubbard model in 1d. This method exploits the identity, $\Xhat^2_{l,l+1}=\Identity$, and enables us to completely separate the spins from the orbitals on a chain with open boundaries by applying a suitably defined non-local unitary transformation on $\Hhat$. The details are as follows.

Consider the $J_\perp$ term in $\Hhat$ which couples the spins with orbitals. Let us denote it as $\Hhat_\perp$.
\begin{equation}
\Hhat_\perp = J_\perp\sum_{l=1}^{L-1}(\tau^+_l \Xhat^{ }_{l,l+1} \tau^-_{l+1} + \tau^+_{l+1} \Xhat^{ }_{l+1,l} \tau^-_{l})
\label{eq:Hperp}
\end{equation}
Here, $\Xhat_{l+1,l}=\Xhat^\dag_{l,l+1}=\Xhat^{ }_{l,l+1}$ (it is Hermitian), and $L$ is the total number of lattice sites. We have explicitly put in the open boundary condition in the summation over $l$. 

Now, on the bond (1,2), define a unitary operator, $\calU^{ }_{1,2}= \Phat^-_2+\Phat^+_2 \Xhat^{ }_{2,1}$, where $\Phat^\pm_2=(1\pm \tau^z_2)/2$ are the projection operators for the orbital states on site 2. By applying $\mathcal{U}_{1,2}$ on $\Hhat_\perp$, we get
\begin{eqnarray}
 &\calU^\dag_{1,2} & \Hhat_\perp \calU^{ }_{1,2}  \nonumber\\
&=& J_\perp \Bigg\{ (\tau^+_1\tau^-_2 + \tau^+_2\tau^-_1 ) + (\tau^+_2 \calXhat^{ }_{2,3} \tau^-_3 + \tau^+_3 \calXhat^{ }_{3,2} \tau^-_2) \nonumber\\
&& + \sum_{l=3}^{L-1} (\tau^+_l \Xhat^{ }_{l,l+1} \tau^-_{l+1} + \tau^+_{l+1} \Xhat^{ }_{l+1,l} \tau^-_{l})\Bigg\}.\end{eqnarray}

As a result of the transformation under $\calU^{ }_{1,2}$, three things have happened to $\Hhat_\perp$. First, the spin-exchange operator $\Xhat^{ }_{1,2}$ has vanished from the bond (1,2). Now we only have $(\tau^+_1\tau^-_2 + \tau^+_2\tau^-_1)$. Second, the operator, $\Xhat^{ }_{2,3}$, on bond (2,3) has been replaced by $\calXhat^{ }_{2,3}=\Xhat^{ }_{1,2}\Xhat^{ }_{2,3}$. Thus, the $\Xhat^{ }_{1,2}$ hasn't quite disappeared from $\Hhat_\perp$. Instead, it has been shifted to the bond (2,3). Unlike $\Xhat^{ }_{2,3}$, the new operator $\calXhat^{ }_{2,3}$ is not Hermitian. But it is unitary, which is sufficient for our method to work. In our notation, $\calXhat^{ }_{3,2}=\calXhat^\dag_{2,3}$. Therefore, $\calXhat^{ }_{2,3}\calXhat^{ }_{3,2} = \calXhat^{ }_{3,2}\calXhat^{ }_{2,3} = \Identity$. And third, the interactions on the bonds beyond the bond (2,3) remain unaffected. These observations suggest that we may similarly transfer $\calXhat^{ }_{2,3}$ to bond (3,4) and so on, and eventually get rid of all the spin-exchange operators in $\Hhat_\perp$ on an open chain. 

Our strategy is to remove the spin-exchange operators from the successive bonds one-by-one. The unitary transformation which does this for us can be defined as: \(\calU = \prod_{l=1}^{L-1} \calU^{ }_{l,l+1}\), where
\begin{equation}
\calU^{ }_{l,l+1}= \Phat^-_{l+1}+\Phat^+_{l+1}\calXhat^{ }_{l+1,l}.
\label{eq:Ul}
\end{equation}
Here, $\calXhat^{ }_{l+1,l} = \Xhat^{ }_{l+1,l} \, \Xhat^{ }_{l,l-1}\cdots\Xhat^{ }_{3,2}\,\Xhat^{ }_{2,1}$
is the string of spin-exchange operators, and $\Phat^\pm_{l+1}=(1\pm\tau^z_{l+1})/2$ are the orbital projectors. Clearly, the $\calU$ is a very non-local unitary operator. By transforming $\Hhat_\perp$ under $\calU$, we get
\begin{equation}
\calU^\dag\Hhat_\perp\calU = J_\perp\sum_{l=1}^{L-1} (\tau^+_l\tau^-_{l+1} + \tau^+_{l+1}\tau^-_l).
\label{eq:UHperpU}
\end{equation}
It is a remarkable transformation which completely decouples the orbitals from the spins. This decoupling happens because the spin-exchange operators that accumulate on the $(L-1,L)$ bond are finally thrown out of the chain, as there is no $(L,1)$ bond on the open chain. It is a beautiful case of exact spin-orbital separation in a model of coupled spins and orbitals. On a closed chain of finite $L$, things complicate. But for a thermodynamically large $L$, the two chains may behave similarly.

The resultant orbital only problem in Eq.~(\ref{eq:UHperpU}) is the exactly solvable XY chain which turns into a free fermion model under the Jordan-Wigner transformation.~\cite{XYmodel} Moreover, the complete absence of spin-spin interaction in the transformed problem makes the spin subsystem an ideal paramagnet. Thus, the $\Hhat_\perp$ is a soluble spin-orbital model, where one exactly knows all the eigenvalues and eigenstates. Every eigenvalue of $\Hhat_\perp$ is exponentially degenerate ($\sim 2^L$) due to the free spins. Even the ground state 
has an extensive entropy of $L\log{2}$. 

We note that $\calU^\dag\tau^z_l\,\calU$ $=$ $\tau^z_l$ and $\calU^\dag\sum_l\sigma^\alpha_l\,\calU$ $=$ $\sum_l \sigma^\alpha_l$ for $\alpha=x,y,z$. Since $\sum_l\vec{\sigma}_l$ and $\tau^z_l$ operators are invariant under $\calU$, it enables us to write a more general model, solvable through the same spin-orbital decoupling procedure as used for the $\Hhat_\perp$ in Eq.~(\ref{eq:UHperpU}).  A simple and realistic modification that we do to $\Hhat_\perp$ is to add the nearest-neighbor orbital Ising term, $J_z\sum_l\tau^z_l\tau^z_{l+1}$. This gives us the model $\Hhat$ of Eq.~(\ref{eq:H}). Clearly, due to the invariance of $\tau^z_l$ under $\calU$, the $\Hhat$ also shows exact spin-orbital separation on an open chain. That is, 
\begin{equation}
\calU^\dag\Hhat\,\calU = \sum_{l=1}^{L-1}\left[J_z\tau^z_l\tau^z_{l+1}+J_\perp (\tau^+_l\tau^-_{l+1} + \tau^+_{l+1}\tau^-_l)\right],
\label{eq:UHU}
\end{equation}
which is the XXZ Heisenberg model, solvable by Bethe ansatz.~\cite{book.sutherland, OpenXXZ, OpenXXZ.sirker, 1dQM} Thus, the $\Hhat$ is a soluble spin-orbital model.
We can in fact add more arbitrary terms of the type, $V_o(\{\tau^z_l\}) + V_s(\sum_l\sigma^z_l,\sum_l\sigma^x_l,\sum_l\sigma^y_l)$, to $\Hhat$, and still rigorously achieve spin-orbital separation under $\calU$. For example, we can certainly add a term like, $\sum_l(h\sigma^z_l+\eta\tau^z_l)$, where $h$ and $\eta$ are the fields for spins and orbitals, respectively. But such more general spin-orbital-separable problems may not always be analytically soluble. Hence, the $\Hhat$ is a special model indeed. 
\subsection{Ground state}
As a function of $J_z$ (for $J_\perp>0$), the exact ground state of $\Hhat$ behaves as follows. For $2|J_z|/J_\perp<1$, the orbital part of the ground state is critical with power law correlations and gapless excitations, akin to the XY case ($J_z=0$). For $2J_z/J_\perp<-1$, the ground state is ferro-orbital and the elementary orbiton excitations are gapped. By ferro-orbital we mean the ferromagnetic state of orbital pseudospins. Moreover, an orbiton is a dispersing orbital excitation, like what a magnon is for a magnet. For $2J_z/J_\perp>1$, the orbital ground state is N\'eel ordered  with gapped orbital excitations. In all these qualitatively different phases, the spins remain completely free (paramagnetic). 

At this point,  we also like to put our understanding of $\Hhat_\perp$ in perspective with some results in Ref.~\onlinecite{Pati.Singh}, where the $\Hhat_{so}$ for $\Delta=\Delta^\prime=0$ has been investigated using DMRG. Particularly for $J_1=0$, the point $J_2/J_3=1$ (which is $J_2=1/4$ in their notation) was identified as a transition point (see Fig.~1 in Ref.~\onlinecite{Pati.Singh} ), above which the ground state is a direct product of the ferromagnetic spins and the orbital fermi-sea (through Jordan-Wigner mapping), $|F\rangle_s\otimes|JW_{fs}\rangle_o$. Here, $F$ stands for  ferromagnetic, $JW_{fs}$ for the Jordan-Wigner fermi-sea, and the subscripts $s$ and $o$ indicate the spin and the orbital subsystems, respectively. For $J_2/J_3\le 1$, they proposed a spin-dimerized antiferromagnetic ($DAF$) ground  state, $|DAF\rangle_s\otimes|JW_{fs}\rangle_o$. Here, we make an important observation that their transition point $J_2/J_3=1$ is same as our $\Hhat_\perp$. In the light of our exact findings, the $J_2/J_3=1$ is a special point in their quantum phase diagram, hitherto unrealized, with exact solvability for the complete eigen-spectrum. Moreover, the correct ground state at this special point is not as stated in Ref.~\onlinecite{Pati.Singh}. Instead, it is a highly entropic manifold of $2^L$ eigenstates, $ \{\calU\,|s_1,s_2,\dots,s_L\rangle\otimes|JW_{fs}\rangle_o\}$, where $|s_1,s_2,\dots,s_L\rangle$ denotes an Ising state in the Hilbert space of $L$ spin-1/2's, with $s_l = \uparrow$ or $\downarrow$ (for $l=1,2,\cdots L$). 

\subsection{Correlations} For the model $\Hhat$ quite a few different correlation functions can be exactly computed,  or understood using the results known for the XXZ chain.

The simplest thing one can compute is the static spin susceptibility, $\chi_s \sim \frac{\beta}{L}\{\langle\left(\sum_l\sigma^z_l\right)^2\rangle - \langle\sum_l\sigma^z_l\rangle^2\}$, where $\beta=1/k_BT$ is the inverse temperature. Consider $\langle\sum_l\sigma^z_l\rangle = tr\left\{\rhohat \sum_l\sigma^z_l\right\}$, where $\rhohat$ is the equilibrium thermal density matrix, $e^{-\beta\Hhat}/Z$, for $\Hhat$. For the decoupled Hamiltonian, $\calU^\dag \Hhat\,\calU$, the thermal density matrix is $\calU^\dag\rhohat\,\calU=\frac{1}{2^L}\Identity\otimes\rhohat^{ }_{o}$, where $\Identity$ is the $2^L$-dimensional identity matrix for the spin subsystem and $\rhohat^{ }_{o}$ is thermal density matrix for the XXZ orbitals. Since $\sum_l\sigma^z_l$ is invariant under $\calU$, the expectation $\langle\sum_l\sigma^z\rangle=\frac{1}{2^L}tr_s \left\{\sum_l\sigma^z\right\}=0$, as it ought to be for a paramagnet. Moreover, $\langle\left(\sum_l\sigma^z\right)^2\rangle =L$. Therefore, $\chi_s\sim 1/{k_BT}$. Here, $tr_s$ denotes the trace over spins only. Similarly, $tr_o$ is the trace over orbitals, and $tr=tr_s\,tr_o$. 

For the orbital pseudospins, the longitudinal susceptibility, $\chi_o\sim \frac{\beta}{L}\{\langle\left(\sum_l\tau^z_l\right)^2\rangle - \langle\sum_l\tau^z_l\rangle^2\}$, can also be calculated from the decoupled problem because the $\tau^z_l$ operators are invariant under $\calU$. The XXZ Heisenberg chain has a very rich mathematical literature from which we can gladly quote the results for $\chi_o$. For the isotropic XXX orbital case, $\chi_o\sim 1+ \frac{1}{2\ln{(T_0/T)}}$ for small temperatures.~\cite{chi_o} In the anisotropic case, for $2J_z/J_\perp>1$, the orbital excitations are gapped. Therefore, $\chi_o$ will exhibit exponential behavior at low temperatures. However, for $2J_z/J_\perp<1$ when the excitations are gapless, through the Bethe ansatz and field-theory treatment, it has been shown that $\chi_o \sim \frac{\theta}{\pi(\pi-\theta)\sin{\theta}} + c T^\zeta$, at low temperatures. Here, $\theta=\cos^{-1}{(2J_z/J_\perp)}$, and $c$ is some constant. The exponent $\zeta= \frac{4\theta}{\pi-\theta}$ for $\theta<\frac{\pi}{3}$, and equal to $2$ otherwise.~\cite{chi_o} For the corrections arising due to the open boundary condition, one may see Ref.~\onlinecite{OpenXXZ.sirker} and the references therein. 

Next we discuss the two-point orbital correlation function, $C^{zz}_o(r) = \langle \tau^z_l\tau^z_{l+r} \rangle$ $=tr\left\{\rhohat\tau^z_l\tau^z_{l+r}\right\}$. The invariance of $\tau^z_l$ operators under $\calU$ implies $C^{zz}_o(r) = tr_o\left\{\rhohat_o\tau^z_l\tau^z_{l+r}\right\}$. The $zz$ orbital correlation is thus exactly same as in the corresponding XXZ problem. For example, in the ground state of $\Hhat_\perp$, it is $C^{zz}_o(r)=-\left[\frac{2}{\pi\,r}\right]^2$ for odd integer values of $r$ and 0 for even $r$.~\cite{XYmodel} For the full problem with non-zero $J_z$, it has been a terribly hard job to find amicable analytic forms of the correlation functions, but using field theoretic techniques, the asymptotic behavior has been predicted to be $C^{zz}_o(r)\sim (-1)^{|r|} \sqrt{\ln{|r|}}/|r|$ for the isotropic XXX case. Unlike the $zz$ correlation function, the computation of $xx$ and $yy$ orbital correlations does not simplify here, because $\tau^{x}_l$ and $\tau^{y}_l$ operators are not-invariant under $\calU$.

The presence of an external magnetic field, $-h\sum_l\sigma^z$, however, simplifies the computation of some ground state correlations by selecting the fully polarized spin state. For $h>0$, the ground state wavefunction is $|\psi_g\rangle=\calU|\uparrow\uparrow\dots\uparrow\rangle_s\otimes|\mbox{XXZ}\rangle_o$ $=|\uparrow\uparrow\dots\uparrow\rangle_s\otimes|\mbox{XXZ}\rangle$, where $|\uparrow\uparrow\dots\uparrow\rangle_s$ and  $|\mbox{XXZ}\rangle_o$ denote the fully polarized spin state and the XXZ orbital ground state, respectively. Note that the $\calU$ acts like an identity operator on the fully polarized spin state. Therefore, the spin-spin correlation in this case is trivial, as all the spins are pointing in the same direction. One also knows the $xx$ and $yy$ orbital correlations in some cases. For example, in the ground state of $\Hhat_\perp$ (with $h\neq 0$), $\langle\psi_g|\tau^x_l\tau^x_{l+r}|\psi_g\rangle = \langle\psi_g|\tau^y_l\tau^y_{l+r}|\psi_g\rangle \sim (-1)^{|r|}/\sqrt{|r|}$, as known for the XY chain in the limit of large $r$.~\cite{XY.McCoy} Surely, one can quote more results for various calculable objects, as the literature for the XXZ model is vast. But we stop it here. Next, we discuss a generalization of the $\Hhat$ for arbitrary orbital pseudospins. Moreover, we also re-look at our model problem in the light of Sr$_2$CuO$_3$.

\section{Miscellaneous remarks}

\subsection{Generalization for orbital pseudospin $\ge$ 1/2} Here, we present a case of spin-orbital separation in systems with more than two orbitals per site. The idea is to demonstrate that, in principle, this phenomenon can also occur when the orbital pseudospin quantum number, $T$, is greater than $\frac{1}{2}$. As an example, consider the model, \[ \Hhat_{T}= \sum_l [J_z \That^z_l\That^z_{l+1} + J_\perp (\That^{+2T}_{l}\Xhat_{l,l+1} \That^{-2T}_{l+1} + h.c.)], \] for an arbitrary $T$. Here, $\That^z_l$ and $\That^\pm_l$ are the angular momentum operators representing the orbital pseudospin on the $l^{th}$ site, and $\That^{\pm 2T}_l = (\That^\pm_l)^{2T}$. We find that $\Hhat_T$ also exhibits complete spin-orbital decoupling under the unitary transformation, $\calU_T$ $=$ $\prod_{l=1}^{L-1} \calU^{ }_{T}(l,l+1)$, where $\calU^{ }_{T}(l,l+1)$ $=$ $\Phat^{-T}_{l+1}+(\mathbf{1}-\Phat^{-T}_{l+1})\calXhat^{ }_{l+1,l}$ is a generalization of Eq.~(\ref{eq:Ul}). Here, $\Phat^{-T}_l$ $=$ $|-T\rangle_l\langle-T|_l$ is the projector for the lowest eigenstate of $\That^z_l$. We can show that $\calU_T^\dag \Hhat_T\calU^{ }_T$ $=$ $\sum_l [J_z \That^z_l\That^z_{l+1}$ $+$ $J_\perp(\That^{+2T}_{l}\That^{-2T}_{l+1} + h.c.)]$, similar to the $\Hhat$. Likewise, it is also valid for the more general forms of $\Hhat_T$, as $\That^z_l$ operators and $\sum_l\vec{\sigma}_l$ are invariant under $\calU_T$. We can also  generalize by allowing for the $\Xhat_{l,l+1}$'s to be unitary (and not only Hermitian) operators, and still have the spin-orbital decoupling by the same method.~\cite{BK.09}


\subsection{Discussion in relation to Sr$_2$CuO$_3$} 
The Sr$_2$CuO$_3$ is a quasi-1d spin-1/2 Heisenberg antiferromagnet, wherein the hole in the $3d^9$ configuration of each Cu$^{2+}$ ion  resides in the $d_{x^2-y^2}$ orbital. Moreover, it has a large energy gap to the orbital excitations ($\sim2.5$ eV from $d_{x^2-y^2}$ to $d_{xz}$). Therefore, it is also a system with ferro-orbital order. Very recently, the Sr$_2$CuO$_3$ has been reported to show spin-orbital separation.~\cite{nature.Sr2CuO3} Given that we also have a model exhibiting spin-orbital separation, it would be interesting to 
draw some connections between our model 
and Sr$_2$CuO$_3$ in the phenomenological spirit. Of course, it is not to suggest that ours is a microscopically derived model for this material.

As the $d_{xz}$ orbital excitation in Sr$_2$CuO$_3$ happens to be most dispersive of them all, we  focus on $d_{x^2-y^2}$ and $d_{xz}$ as the basis for the two-level orbital pseudospin. Since the two orbitals are separated by a large (crystal field) energy, $\eta$, that selects the ferro-orbital state, it requires us to have the term, $\frac{\eta}{2}\sum_l\tau^z_l$, in the Hamiltonian. For the spin part, we take nearest-neighbor antiferromagnetic exchange interaction. We model the coupling between spins and orbitals by $\Hhat_\perp$ of Eq.~(\ref{eq:Hperp}), a personally favored spin-orbital-separable choice. Thus, a minimal phenomenological model relevant to Sr$_2$CuO$_3$ could be, 
\begin{equation}
\Hhat_\perp + \frac{J}{4}\sum_l\vec{\sigma}_l\cdot\vec{\sigma}_{l+1}+\frac{\eta}{2}\sum_l\tau^z_l,
\label{eq:sr2cuo3}
\end{equation}
 with $J\sim .25$eV and $\eta \sim 2.5$eV. The $J_\perp$ can be estimated from 
 the orbiton energies (bandwidth $\sim .2$eV). Here, we may also include the term, $J_z\sum_l\tau^z_l\tau^z_{l+1}$, but it doesn't seem necessary. 
 
 Notably, the $J$ term here presents a `formal' difficulty to rigorous spin-orbital separation, as it is not invariant under $\calU$. But the above model has in it the basic features of Sr$_2$CuO$_3$.
 For a strong positive $\eta$, the exact ground state of this model is obviously ferro-orbital, and it is described by the antiferromagnetic Heisenberg model for the spin part, as is the case for Sr$_2$CuO$_3$. It can also be checked that, through $\Hhat_\perp$, an excited state created by a local orbital `flip' in an antiferromagnetic spin background evolves into separately dispersing orbitons and spinons [as sketched in Fig. 1(a) of Ref.~\onlinecite{nature.Sr2CuO3}]. This looks the same as discussed recently in the SU(4) model in an orbital field (using a different approach than ours),~\cite{Wohl.Brink} and was used to describe Sr$_2$CuO$_3$. Hence, it appears that the model of Eq.~(\ref{eq:sr2cuo3}) could likewise be useful.

This discussion further motivates us to write down models which have antiferromangetic spin ground states and show exact spin-orbital separation under $\calU$. To do this, we replace $\sum_l\vec{\sigma}_l\cdot\vec{\sigma}_{l+1}$ in Eq.~(\ref{eq:sr2cuo3}) by $\left(\sum_l\vec{\sigma}_l\right)^2$, or by $\Phat^-_o\sum_l\vec{\sigma}_l\cdot\vec{\sigma}_{l+1}\Phat^-_o$,  where $\Phat^-_{o}=\prod_l(1-\tau^z_l)/2$ is a ferro-orbital projector. Since $\calU$ does not affect these new terms, the two cases show exact spin-orbital separation under $\calU$, while their ground states are spin-singlets.

\section{Summary} We have found a solvable 1d spin-orbital model, $\Hhat$ [of Eq.~(\ref{eq:H})]. It shows exact spin-orbital separation under the unitary transformation, $\calU$, which systematically rids the $\Hhat$ of the spins that are coupled to the orbital fluctuations. The transformed problem has free spins and the XXZ model for orbitals. While the obvious symmetry, SU(2)$\times$U(1), of $\Hhat$ implies the conservation of $\sum_l\vec{\sigma}_l$ and $\sum_l\tau^z_l$ only, in actual, all the spins that vanish from $\Hhat$ under $\calU$ are conserved. This fact and the integrability of the XXZ chain imply that we have a complete knowledge of all the conserved quantities of $\Hhat$. We have also presented a generalization of $\Hhat$ for the orbital pseudospins $>1/2$, exhibiting spin-orbital separation.

In view of the recent experimental observations of spin-orbital separation in Sr$_2$CuO$_3$, we have briefly remarked on the scope of our model in relation to Sr$_2$CuO$_3$. While we seem to have the minimal ingredients [as in Eq.~(\ref{eq:sr2cuo3})] required to discuss spin-orbital separation in Sr$_2$CuO$_3$, a proper comparison needs explicit calculations. 

A point of further study in our models would be to understand the nature of spin-orbital entanglement.~\cite{Oles.Entangle} Since the spins and orbitals completely decouple here (in the $\calU$-transformed basis), one might think of it as having no spin-orbital entanglement. But remember that $\calU$ is a very non-local transformation. Therefore, in the original basis, the different eigenstates of $\Hhat$ may actually have non-zero entanglement up to different degrees.  
After all, in general, the entanglement is not invariant under the unitary rotations of the full system. We will discuss this in detail elsewhere.

\begin{acknowledgments}
The author thanks Frederic Mila for useful comments and encouragement.
The partial financial support from the DST-PURSE is gratefully acknowledged.
\end{acknowledgments}

\bibliography{manuscript.bib}
\end{document}